\documentclass[%
superscriptaddress,
preprint,
amsmath,amssymb,
 ps,
showpacs,
prb,
]{revtex4-2}

\usepackage{graphicx}
\usepackage{dcolumn}
\usepackage{bm}

\begin{document}


\title{Coexistence of parallel and series current paths in parallel-coupled double quantum dots in nonlinear transport regime
}

\author{Tsuyoshi Hatano}%
\affiliation{College of Engineering, Nihon University, Koriyama, Fukushima, 963-8642, Japan}

\author{Toshihiro Kubo}
\affiliation{National Institute of Technology, Tsuyama College, Tsuyama, Okayama, 708-8509, Japann}

\author{Shinichi Amaha}
\affiliation{Center for Emergent Matter Science, RIKEN, Wako-shi, Saitama, 351-0198, Japan}

\author{Yasuhiro Tokura}
\affiliation{Faculty of Pure and Applied Sciences, University of  Tsukuba, Tsukuba 305-8571, Japan}
\affiliation{Tsukuba Research Center for Energy Materials Science (TREMS), Tsukuba 305-8571, Japan}

\author{Seigo Tarucha}
\affiliation{Center for Emergent Matter Science, RIKEN, Wako-shi, Saitama, 351-0198, Japan}

\date{\today}

\begin{abstract}
We investigated the electron transport properties of parallel-coupled double quantum dot (DQD) devices under magnetic fields. 
When a low magnetic field was applied, electron tunneling through parallel-coupled DQDs was observed. 
Under a high magnetic field, we observed both electron tunneling through parallel- and series-coupled DQDs under nonlinear transport conditions. 
In addition, the Pauli spin blockade was observed, indicating tunneling through the series-coupled DQDs.
We attribute these behaviors to the magnetic-field-induced changes in the tunnel-couplings that allow the coexistence  of the current paths of the parallel and series configurations.
\end{abstract}

\maketitle

Quantum dots (QDs) are nanoscale devices in which electron charges and spins can be controlled one by one. 
Therefore, QD devices are expected to be applied to not only electronics but also spintronics. 
Furthermore, electron spins in QDs are employed as quantum bits (qubits), information units of quantum computing.

Double QDs (DQDs) are formed by coupling two QDs quantum mechanically and electrostatically. 
There are two types of DQDs: series- and parallel-coupled DQDs. 
For series-coupled DQDs, there is only one current path through which electrons are transported between two electrodes through two coupled QDs in series, and current suppression due to the Pauli spin blockade has been observed.\cite{Ono,Amaha1,Kondo}. 
Furthermore, qubit operation has been demonstrated by observations that the Pauli spin blockade is lifted by coherent spin rotation.
\cite{Koppens,Laird,Nowack,Pioro,Petta,Brunner,Noiri,Ono2}.  
In contrast, for parallel-coupled DQDs, there are multiple current paths 
because electrons are transported between two electrodes through not only the two QDs but also one  QD. 
Therefore, ground and excited states, \cite{Hatano1,Hatano2} as well as Aharonov-Bohm (AB) oscillations, \cite{Hatano3} have been observed.  
Consequently, the observed electronic transport properties of the series- 
and parallel-coupled DQDs are different, 
and the two transport properties have not been observed in a single DQD device.

In this study, we investigated the electron transport properties of parallel-coupled DQD devices 
under a magnetic field. 
At low magnetic fields, only the electronic properties of the parallel-coupled DQDs were observed. 
On the other hand, applying a high magnetic field, both features of parallel- and series-coupled DQDs were observed  
in the nonlinear transport regime. 
Furthermore, the Pauli spin blockade was also observed, indicating tunneling through the series-coupled DQDs. 

\begin{figure}[h]
\includegraphics[width=0.8\columnwidth]{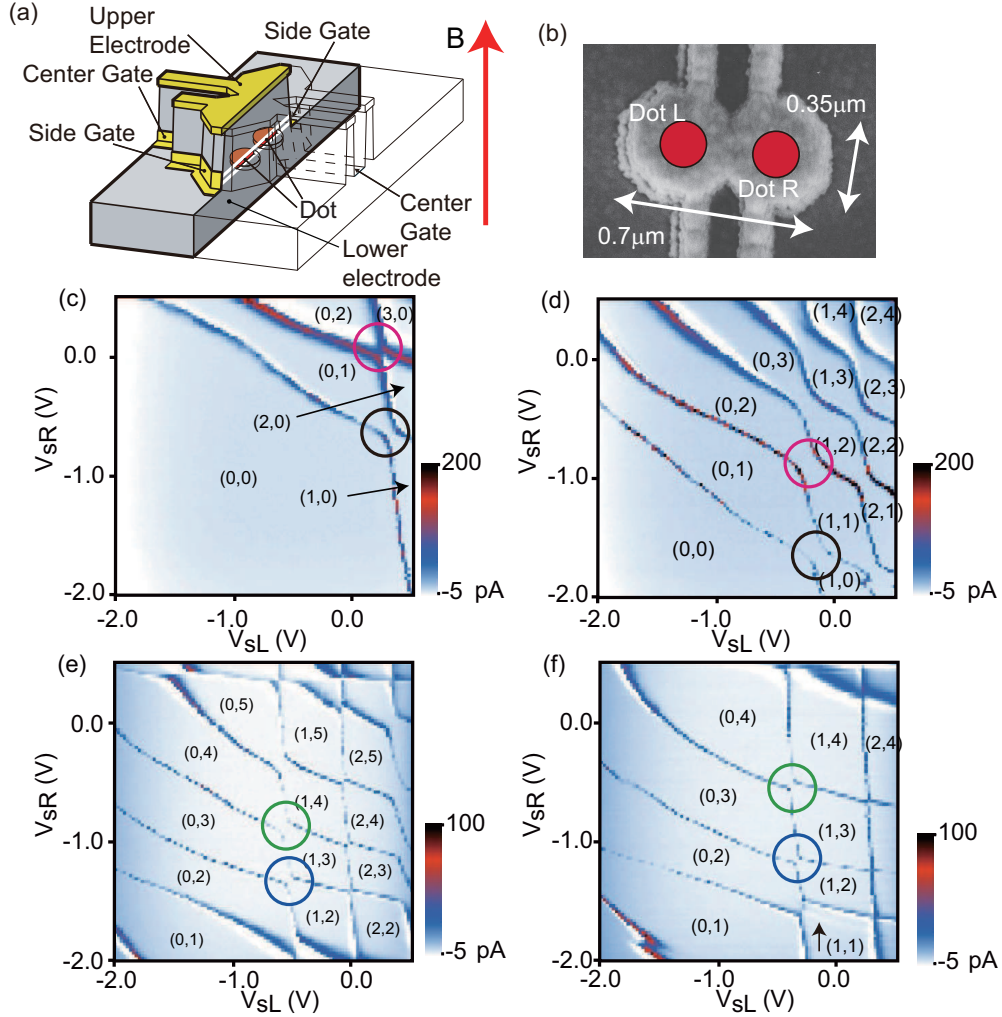}
\caption{
(a) Schematic of the structure of the parallel-coupled DQD device.  
The red arrow indicates the direction of the magnetic field.
(b) Scanning electron microscope image of the parallel-coupled DQD device.  
The false red circles indicate the position of the QDs.
(c-f) Color plots of current $I$ as a function of  the left and right side gate voltages $V_{sL}$ 
and $V_{sR}$ (stability diagrams) at the bias voltage $V_{b}=0.1$ mV, the magnetic field $B=0$  at the center gate voltage (c)$V_c=-1.6$ V and (d) $V_c=-1.3$ V.  
Stability diagram at $V_c=-0.7$ V, $V_{b}=10$ $\mu$V and (e)$B=8$ T, and (f) $B=12$ T. 
$(N_L, N_R)$ indicates the electron numbers of dots L and R, respectively. 
} 
\label{device}
\end{figure}

The parallel-coupled DQD device is shown in Fig.~\ref{device} (a). 
It is made from a double-barrier heterostructure (DBH) 
and consists of two laterally coupled vertical QDs with four split gates. 
The DBH consists of an undoped 10-nm GaAs well and undoped 8 and 7-nm-thick AlGaAs barriers 
(the thicker one is closest to the substrate). 
Two of the gates (side gates) are used to independently tune the electron number in each QD, 
and the remaining two gates (center gates) are used to tune the interdot tunnel coupling\cite{Hatano1}. 
A magnetic field $B$ is applied perpendicularly to the substrate. 
The scanning electron micrograph of the device is shown in Fig.~\ref{device} (b). 
The dots L and R, indicated by two red circles,  
are located inside the $0.35\times 0.7$ $\mu$m$^2$. Four line mesas emerge from the sides of the DQD mesa. 
The mesas are sufficiently thin that current flows only through the top metal contact. 
The line mesas split the surrounding Schottky gate metals \cite{Austing,Hatano4}. 
Transport measurements were conducted in a dilution refrigerator at a temperature of 20 mK.

Figure \ref{device} (c) shows the color plot of the current $I$ 
as a function of the left and right side gate voltages  $V_{sL}$ and $V_{sR}$ (stability diagram)
at the center gate voltage $V_c=-1.6$ V, the bias voltage $V_{b}=0.1$ mV, 
and $B=0$ T.  
$(N_L, N_R)$ indicates the electron numbers of dots L and R, respectively. 
Here, the measurements were limited to $-2$ and  $0.5$ V 
because a large leak current flowed through the DQD device outside of this range. 
Figure \ref{device} (c) is a typical stability diagram of the parallel-coupled DQDs 
in the linear transport regime \cite{Hatano1}. 
The regions of no electrons in each dot are clearly visible \cite{Hatano1}.   
Figure ~\ref{device} (d) shows the stability diagram at  $V_c=-1.3$ V. 
The values of $B$ and $V_{b}$ are those in Fig.~\ref{device} (c). 
As $V_c$ becomes high, the minimum separation of the nearest two Coulomb
peaks in the black and red circles in Figs.~\ref{device} (c) and (d) increases,  
indicating the strengthening of the interdot tunnel couplings \cite{Hatano5}. 
Moreover, the position of the Coulomb oscillation peaks shifts toward positive $V_{sL}$ and $V_{sR}$
at high $V_c$.

\begin{figure}[htb]
\includegraphics[width=1\columnwidth]{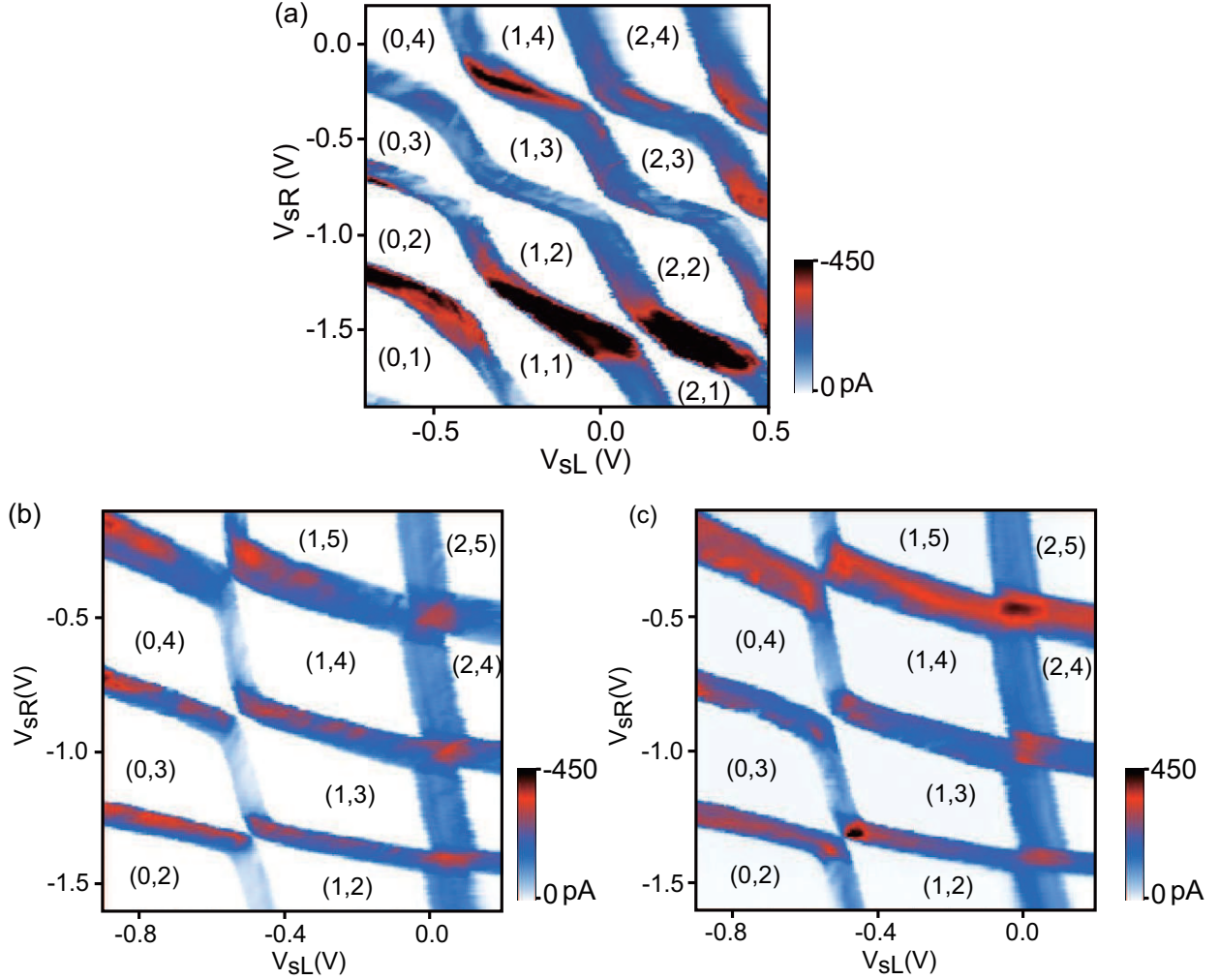}
\caption{
(a) Stability diagram at $V_c=-1.1$ V, $V_{b}=-0.8$ mV, and $B=0$ T.  
Stability diagrams at $V_c=-0.7$ V,   
  $B=8$ T, (b) $V_{b}=-0.6$ mV, and (c) $V_{b}=0.6$ mV. 
} 
\label{currentb8tn3}
\end{figure}

Figure~\ref{device} (e) shows the stability diagram at $V_c=-0.7$ V, $V_{b}=10$ $\mu$V and $B=8$ T. 
The interdot tunnel coupling was weaker than in Figs.~\ref{device} (c) and (d) despite the higher $V_c$ 
because  the out-of-plane magnetic field narrows the spread of the wave functions 
in the two QDs \cite{Hatano4}.     
Figure~\ref{device} (f) shows  the stability diagram at $B=12$ T. 
The values of $V_c$ and $V_{b}$ are those in Fig.~\ref{device} (e). 
In this diagram, the anticrossing occurring at $V_{sL}\sim -1.9$ V and $V_{sR}\sim -1.6$ V 
 and the Coulomb oscillation indicated by the black arrow
are attributable to dopants or interface roughness. 
Comparing panels (e) and (f) of Fig.~\ref{device}, one observes that increasing $B$ weakens 
the interdot tunnel couplings (blue and green circles) and shifts   
 the position of the Coulomb oscillation peaks to positive $V_{sL}$ and $V_{sR}$. 

Figure \ref{currentb8tn3} (a) shows the stability diagram  
at $V_c=-1.1$ V, $V_{b}=-0.8$ mV, and $B=0$ T.  
The thick lines indicate the current-flow regions. 
This stability diagram is a typical one of the parallel-coupled DQDs under nonlinear transport conditions \cite{Hatano1,Hatano2}. 

Figure~\ref{currentb8tn3} (b) shows the stability diagram 
at  $V_c=-0.7$ V,  $V_{b}=-0.6$ mV, and $B=8$ T. 
The values of  $V_c$ and $B$ are the same in Fig.~\ref{device} (e). 
The features of this stability diagram are similar to those in Fig.~\ref{currentb8tn3} (a). 
However, the interdot tunnel couplings 
are weaker than those  in Fig.~\ref{currentb8tn3} (a) despite a higher $V_c$ 
due to the out-of-plane magnetic field \cite{Hatano4}.    
Figure \ref{currentb8tn3} (c) shows  the stability diagram   
at the same value of $V_{b}$ but opposite polarity in Fig.~\ref{currentb8tn3} (b).
The values of $V_c$ and $B$ are the same as those in Fig.~\ref{currentb8tn3} (b). 
The stability diagrams in Figs.~\ref{currentb8tn3} (b) and (c) are almost the same. 
The stability diagram being independent of the polarity $V_b$ is the feature of the parallel-coupled DQDs.

\begin{figure}[htb]
\includegraphics[width=1\columnwidth]{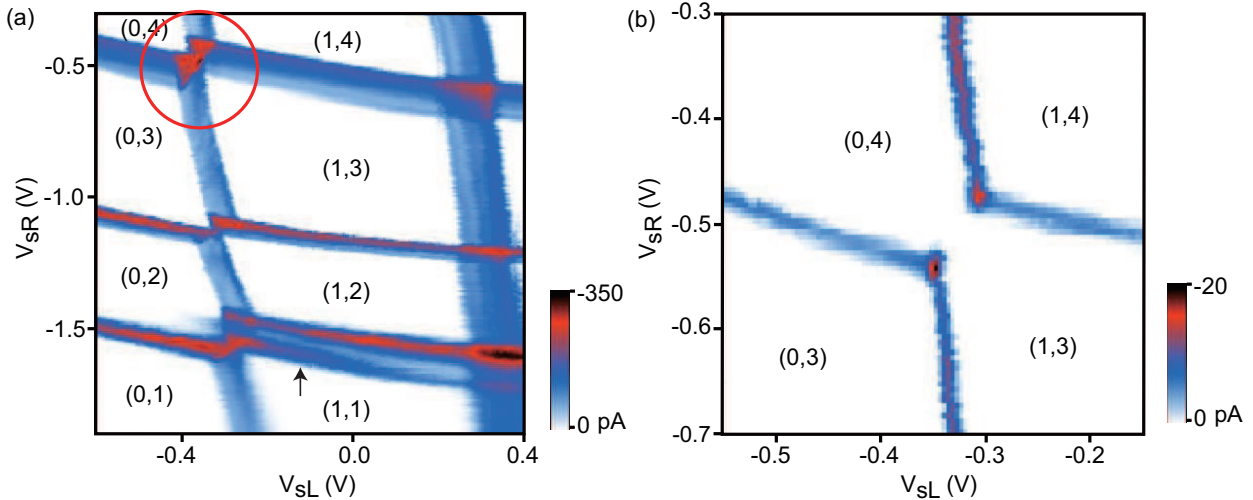}
\caption{
(a) Stability diagram at $V_c=-0.7$ V, $V_{b}=-0.5$ mV, and  $B=12$ T. 
(b) Stability diagram at $V_c=-0.7$ V, $V_{b}=-20$ $\mu$V, and $B=12$ T. 
} 
\label{currentb12t}
\end{figure}

We applied a much higher magnetic field to the parallel-coupled DQD devices. 
The stability diagram obtained at $B=12$ T is shown in Fig.~\ref{currentb12t} (a).  
Here,  $V_{c}=-0.7$ V and $V_{b}=-0.5$ mV. 
As explained above, the Coulomb oscillation peak indicated by the black arrow
is attributable to dopants or interface roughness.	
The interdot tunnel couplings are even weaker than those at $B=8$ T, and 
the stability diagram reflects the feature of that when there is almost only electrostatic coupling \cite{Wiel}.  
However, in the red circle in Fig.~\ref{currentb12t} (a), 
the triangle regions of enhanced current flow, which indicate the feature of the series-coupled DQDs, are observed 
even in the parallel-coupled DQD device.  
Figure~\ref{currentb12t} (b) shows the stability diagram in the linear transport regime ($V_{b}=-20$ $\mu$V) 
in the same region as the red circle in Fig.~\ref{currentb12t} (a).   
The feature of this stability diagram indicates that of the parallel-coupled 
DQDs with weak interdot tunnel coupling in the linear transport regime \cite{Wiel}.

\begin{figure}
\includegraphics[width=1\columnwidth]{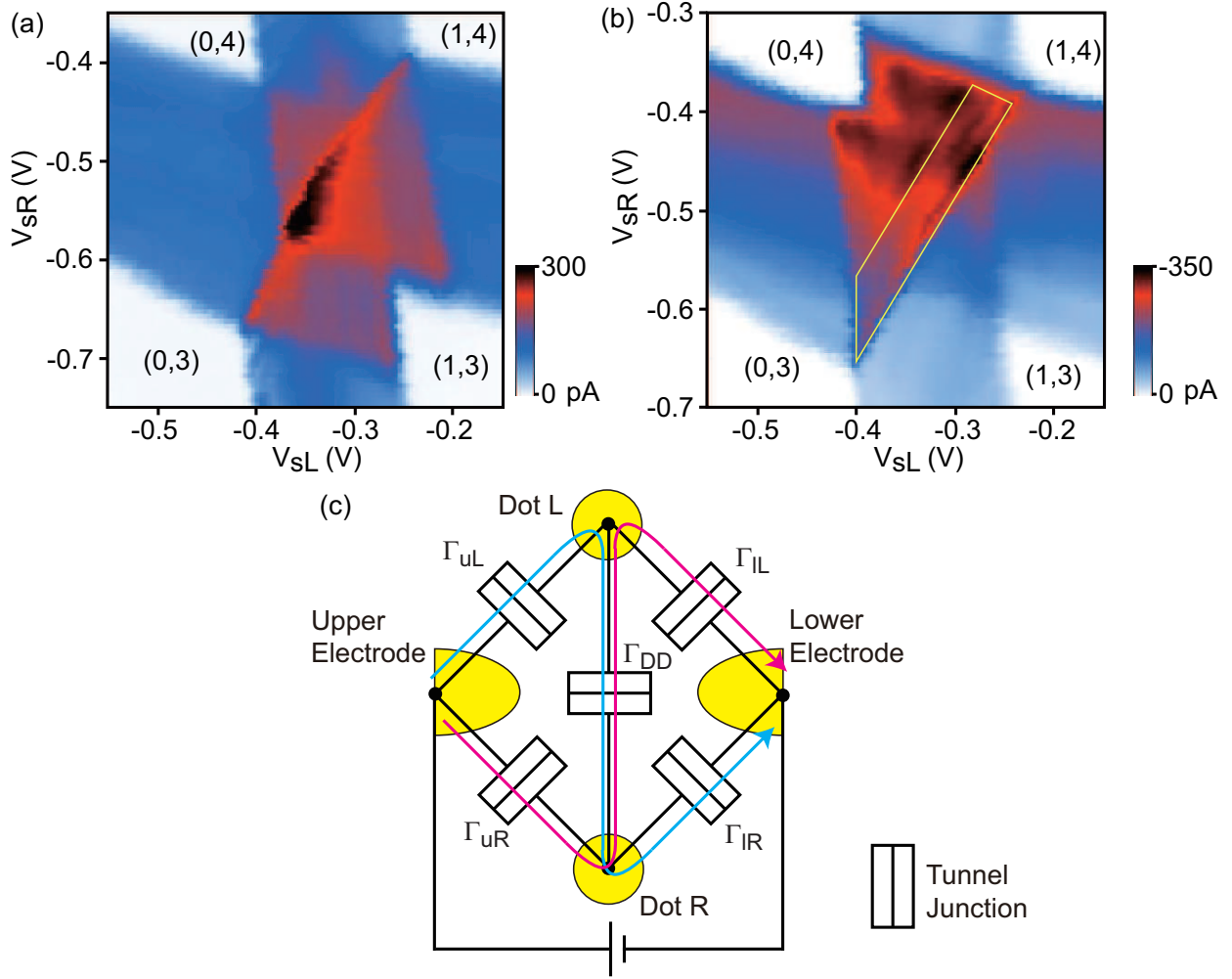}
\caption{
Stability diagrams at $B=12$ T, $V_c=-0.7$ V, (a) $V_{b}=0.8$ mV 
and  (b) $V_{b}=-0.8$ mV. 
(c) Electronic circuit of the parallel-coupled DQD device.  
When the interdot tunnel coupling is sequential, 
the current flows in the direction of the light blue or red arrows. 
The two yellow circles and the two yellow semiellipsoids indicate 
the positions of the two QDs and two electrodes, respectively. 
} 
\label{sbb12t}
\end{figure}

To examine the stability diagram in  the red circle in Fig.~\ref{currentb12t} (a) more closely, 
we obtained the stability diagrams at $B=12$ T and higher  $V_{b}$ of $0.8$ and $-0.8$ mV, and the results 
are shown in Figs.~\ref{sbb12t} (a) and (b), respectively.  
Here, the value of $V_c$ is the same as that in Fig.~\ref{currentb12t} (a). 
Comparing the two stability diagrams, 
the orientation of the triangles at the triple degenerate points is reversed by changing the polarity of $V_{b}$.  
This feature is typically observed in the stability diagram of the series-coupled DQDs 
in the nonlinear transport regime \cite{Wiel}. 
However, different from the series-coupled DQDs, 
thick lines, which indicate that current flows through only one QD, 
are observed. 
This is the feature of the parallel-coupled DQDs.

We consider why the stability diagram obtained at a high magnetic field 
showed the features of the series- and parallel-coupled DQDs simultaneously. 
When zero or low magnetic fields are applied to the parallel-coupled DQD devices, 
the interdot tunnel coupling remains strong.   
Therefore, the tunnel-coupled states of the two QDs, e.g., bonding and antibonding states, are formed\cite{Hatano1,Hatano2}. 
In this case, a typical stability diagram of the parallel-coupled DQDs is observed. 
However, when the magnetic field is high, 
the interdot tunnel coupling becomes very weak.  
Moreover, zero- and one-dimensional emitter states form in the two electrodes    
because near the potential barriers 
of the GaAs/AlGaAs heterostructure (Fig.~\ref{device} (a)), 
the electrodes are about a few hundred nm in size\cite{Leadbeater,Su,Kamata,NIshi}. 
Accordingly, the tunnel coupling between  the two electrodes and dot L and R  are independent of each other 
and have different values because random potentials caused by dopants or interface roughness exist. 
Therefore, the electron tunneling for not only between
the QDs and electrodes but also between two QDs is a sequential process, and hence the current via coherent
superposed states formed in the coupled QDs is negligible. 
In this situation, the potential barriers, which form the QDs, serve as tunnel junctions 
and the electronic circuit of the parallel-coupled DQDs is shown in Fig.~\ref{sbb12t}(c).  
As shown in Supplementary Data, when finite energy difference of energy levels of dots L and R is
induced by $V_{sL}$ and $V_{sR}$ under the condition of different tunnel couplings between two electrodes and 
dots L and R, current paths in the series-coupled DQDs are formed, 
as shown by the red and light blue arrows in Fig.~\ref{sbb12t}(c), 
which is similar to the series-coupled DQDs. 
Note that we can observe the triangles of the current flow 
in the region where the electron numbers in dot L and R change from  $(1,5)$ to $(2,4)$ in Fig.~\ref{currentb8tn3} (b).  
However, the triangle of current flow in the same region is not observed in Fig.~\ref{currentb8tn3} (c). 
Therefore, whether we observe the current path of the series-coupled DQDs
in parallel-coupled DQDs depends on the strength of the tunnel coupling between the QDs.

\begin{figure}
\includegraphics[width=1\columnwidth]{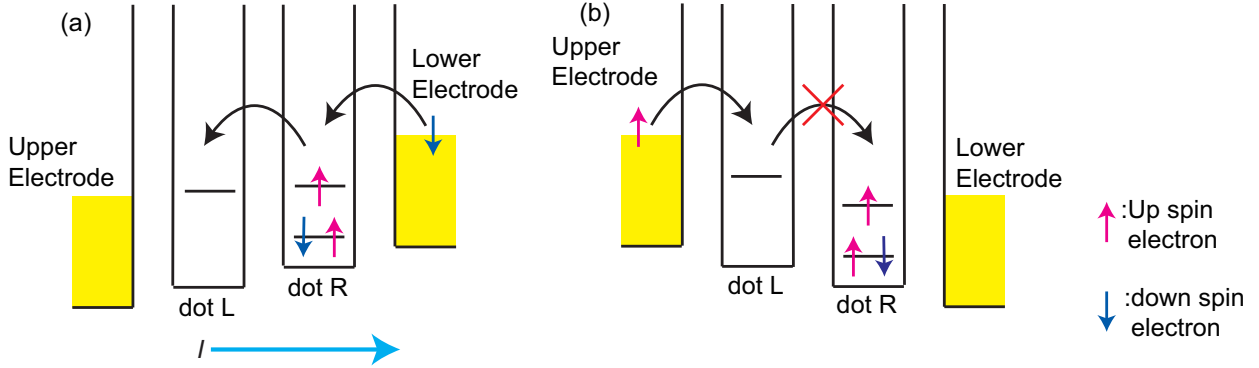}
\caption{
Schematic energy diagrams for (a) $V_{b}>0$ and (b) $V_{b}<0$. 
The red and blue arrows indicate the electrons with the up and down spin, respectively. 
The light blue arrow indicates the direction of the current flow. 
} 
\label{schqbc}
\end{figure}

The variation in the stability diagrams with the polarity of $V_{b}$ was then investigated in more detail 
(Fig.~\ref{sbb12t} (a) and (b)). 
In Fig.~\ref{sbb12t}(a), the triangles of current flow are clearly observed. 
Figure~\ref{schqbc} (a) shows the schematic energy diagram at $V_{b}>0$.  
The up and down arrows indicate the electrons with up and down spins, respectively. 
In general, QDs with a circular harmonic potential are formed in the vertical single QDs 
and vertically coupled DQDs \cite{Tarucha,Amaha2}. 
However, in laterally coupled vertical DQDs (Fig.~\ref{device} (a)), the potential structures of the two QDs 
 are non-circular due to the in-plane interdot tunnel coupling.  
Therefore, the energy levels of each QD are not degenerate and
two electrons with antiparallel spins occupy the energy levels in QDs, 
from the lowest energy level upword.\cite{Hatano1}.  
In Fig.~\ref{device} (f), the values of $B$ and $V_{c}$ are those in Figs.~\ref{sbb12t} (a) and (b), 
and the $(0,4)$ region is wider than the $(1,3)$ region. 
This result indicates that a core closed shell is formed in the $(0,4)$ region of dot R. 
Accordingly, we assume that
an electron with up spin is localized in the $(0,3)$ region of dot R.  
Subsequently, when a positive $V_{b}$ is applied to the DQD devices, 
a down-spin electron can transfer from the lower to the upper electrodes in Fig.~\ref{schqbc} (a) 
and the current flows in the direction of the light blue arrow in Fig.~\ref{schqbc} (a).

At $V_{b}<0$, the direction of the current triangles in Fig.~\ref{sbb12t} (b) opposes that 
in Fig.~\ref{sbb12t} (a)\cite{Wiel}. 
Figure~\ref{schqbc} (b) shows the energy diagram at $V_{b}<0$.  
In Fig.~\ref{schqbc} (b), after an electron with up spin tunnels from the upper electrode to dot L, 
its tunneling from dot L to dot R is prohibited by the Pauli spin blockade imposed 
by the Pauli exclusion principle \cite{Ono}. 
Therefore, in the yellow trapezoid in Fig.~\ref{sbb12t} (b), a part of  the current-flow triangles is missing.
However, the Pauli spin blockade is not  perfect 
because the up-spin electron in dot L directly  
tunnels to the lower electrode via the tunnel coupling between dot L and the lower electrode.   
Accordingly, the Pauli spin blockade  is not observed clearly.

Can we confirm the path of the current flow in Fig.~\ref{sbb12t} (c)? 
The current path can indeed be confirmed using the Pauli spin blockade principle.  
To show this, we constructed a measurement system 
in which the current $I$ is positive at $V_{b}>0$ . 
In the DQD device, the current then flows at $V_{b}>0$ and the Pauli spin blockade occurs at $V_{b}<0$. 
From these measured results, we confirmed 
that the constitution of the two electrodes and the two QDs are as shown in Figs.~\ref{schqbc}(a) and (b). 
Therefore, the current path shown by the light blue arrow in Fig.~\ref{sbb12t} (c) is formed.

The electric circuit of the parallel-coupled DQDs in Fig.~\ref{sbb12t}(c)  is equivalent to a classical bridge circuit.
Therefore, parallel-coupled DQDs, in which all tunnel couplings are sequential processes, can be considered as quantum bridge circuits. 
Here, the line-widths between the lower (upper) electrodes and dots L and R are denoted 
as  $\Gamma_{\mathrm{lL}} (\Gamma_{\mathrm{uL}})$ and $\Gamma_{\mathrm{lR}} (\Gamma_{\mathrm{uR}})$, respectively, 
and the interdot transition rate is represented 
by  $\Gamma_{\mathrm{DD}}$. 
Using the master equation method, we calculated the current flows $I|_{\Gamma_{\mathrm{DD}}=0}$ and $I|_{\Gamma_{\mathrm{DD}}\neq0}$ 
at $\Gamma_{\mathrm{DD}}=0$ and  $\Gamma_{\mathrm{DD}}\neq 0$, respectively (see Supplementary Data). 
The series current $I_{\mathrm{series}}$ flowing through the two QDs is given by 
\begin{eqnarray}
I_{\mathrm{series}}&=&I|_{\Gamma_{\mathrm{DD}}\neq0}-I|_{\Gamma_{\mathrm{DD}}=0}\nonumber\\
&\propto&(\Gamma_{\mathrm{uR}}\Gamma_{\mathrm{lL}}
-\Gamma_{\mathrm{uL}}\Gamma_{\mathrm{lR}})\Gamma_{\mathrm{DD}}
\nonumber. 
\end{eqnarray}
This current expression is similar to that of a Wheatstone bridge.  
In a Wheatstone bridge, 
the value of the unknown resistance can be determined from the values of the three known resistances. 
In contrast, individually estimating  the tunnel couplings of the potential barriers
is difficult in our case. 
Furthermore, the tunnel couplings in the DQDs are difficult to hand-tune 
because cross-talk occurs among  the gate voltages. 
However, if the interdot tunnel coupling could be tuned using computer-automated algorithms \cite{Diepen},  
we might reduce the triangles of current flow to zero by tuning the gate voltages, 
thereby achieving 
$\Gamma_{\mathrm{uR}}\Gamma_{\mathrm{lL}}=\Gamma_{\mathrm{uL}}\Gamma_{\mathrm{lR}}$. 
In this way, we could form the two QD systems with the same tunnel coupling ratio between the QD and two electrodes 
$\Gamma_{\mathrm{uL}}/\Gamma_{\mathrm{lL}}=\Gamma_{\mathrm{uR}}/\Gamma_{\mathrm{lR}}$.


In summary, we examined the electron transport properties of parallel-coupled DQD devices 
under a perpendicular magnetic field. 
For zero or low magnetic fields, the stability diagrams show only properties of parallel-coupled DQDs. 
In contrast, at high magnetic fields, the stability diagrams showed the features of  
the series- and parallel-coupled DQDs in a nonlinear transport regime 
because a finite difference between the energy levels of the two QDs is
induced by side gate voltages when the tunnel couplings differ between the two electrodes and two QDs. 
Furthermore, we observed the Pauli spin blockade in the stability diagram of the parallel-coupled DQD devices.

\begin{acknowledgments}
Part of this work was financially supported by the Japan Science and Technology Agency, JST.  
T. H. was supported by  Nihon University Multidisciplinary Research Grant for (2021). 
Y. T. acknowledges support from JST [Moonshot R\&D][Grant Number JPMJMS2061].
S. T. acknowledges financial support from Core Research for Evolutional Science and Technology (CREST), 
Japan Science and Technology Agency (JST) (JPMJCR15N2 and JPMJCR1675).
\end{acknowledgments}

\pagebreak

\begin{center}
  \textbf{\large Supplementary Data for 
Coexistence of parallel and series current paths in parallel-coupled double quantum dots in nonlinear transport regime}\\[.2cm]
  Tsuyoshi Hatano,  Toshihiro Kubo, Shinichi Amaha, Yasuhiro Tokura and Seigo Tarucha\\[.1cm]
\end{center}

\vspace{2cm}

\def\**{\stackrel{\textstyle *}{*}}
\newcommand{\defe}{\stackrel{\mathrm{def}}{=}}
\newcommand{\we}{\stackrel{\mathrm{w}}{=}}

\setcounter{equation}{0}
\setcounter{figure}{0}
\renewcommand{\thesection}{\large\arabic{section}}
\renewcommand{\thesubsection}{\arabic{section}.\arabic{subsection}}

\section{Introduction}

 To clarify the condition of current triangles in a stability diagram under a finite bias condition, 
we consider sequential tunneling not only between quantum dots (QDs) and
 electrodes but also between two QDs. 
The current via coherent 
 superposed states formed in the coupled QDs is negligible.
 Instead, we assume an incoherent, 
 possibly phonon mediated, interdot tunneling process. 
The current is estimated under a large bias limit using the master equation
 formalism.
 
\begin{figure}[h]
\begin{center}\leavevmode
\includegraphics[width=0.5\linewidth]{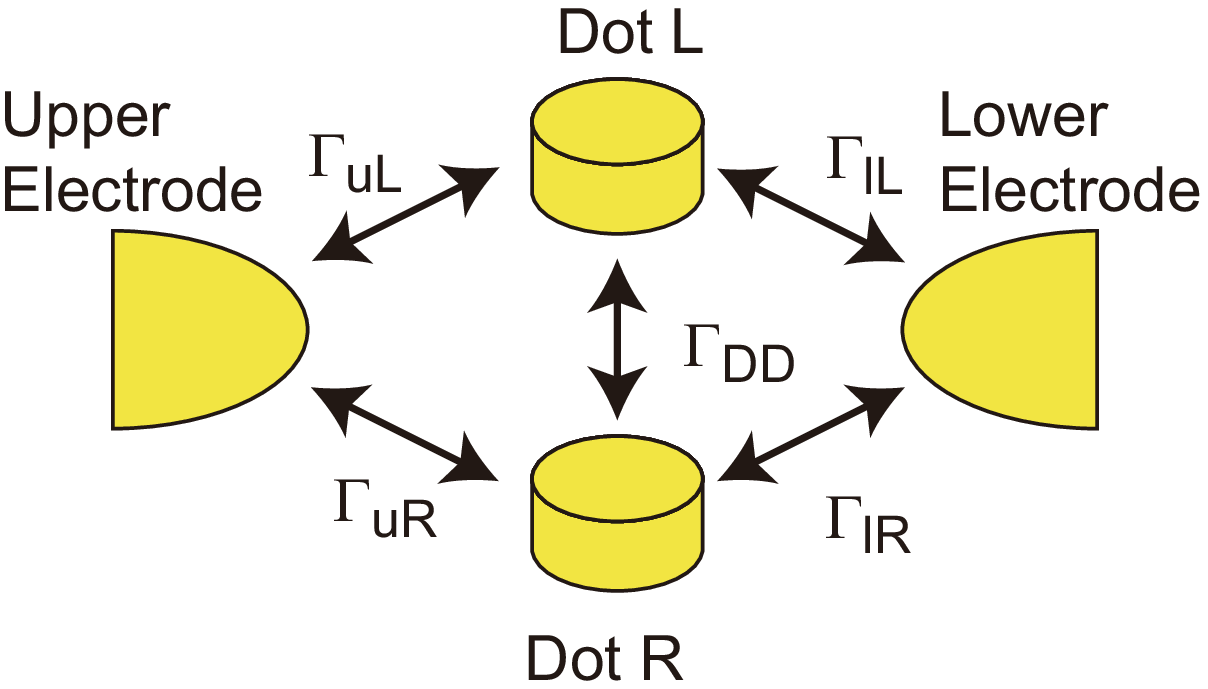}
\caption{Schematic of the parallel coupled double quantum 
dot system.} 
\label{fig:model}
\end{center}
\end{figure}

A model of the system is depicted in Fig.~\ref{fig:model}.
$\Gamma_{\mathrm{lL}}$ $(\Gamma_{\mathrm{uL}})$ 
and $\Gamma_{\mathrm{lR}}$ $(\Gamma_{\mathrm{uR}})$ are
 the line-widths between dots L and R and the lower (upper) electrode, respectively. 
The energy dependence of the line-widths (wide-band limit) is neglected. 
The interdot transition rate $\Gamma_{\mathrm{DD}}$ between the quantum states of dots L and R
 is much larger than $2t_c$, where
 $t_c$ indicates the coherent tunnel coupling energy.
 At low temperatures, 
the rate is finite only from the dot with larger eigenenergy to the dot with smaller eigenenergy.
 For simplicity, we also neglected the dependence of $\Gamma_{\mathrm{DD}}$ 
 on eigenenergy difference between dots L and R.
 We consider that
 only one eigenenergy (including the spin degree of freedom)
 for each QD is available between the bias windows separating the Fermi energies of the two electrodes.

 \section{Independent current path}
 To explain our theoretical method, we first set  
 $\Gamma_{\mathrm{DD}}=0$. Under this condition, namely that the two QDs constitute
 independent current paths.
 This situation also describes the Pauli spin blockade condition in which 
 transitions between the QDs are prohibited by the Pauli exclusion principle.
 As the dynamics of each QD are independent, we focus on QD $\nu=$L or R.
The probability vector of QD $\nu$ at time $t$ is defined as
\begin{align}
\vec{\rho}_{\nu}(t)&=\left(
  \begin{array}{c}
    \rho_{\nu 0}(t) \\
    \rho_{\nu 1}(t)
  \end{array}
\right),
\end{align}
where $\rho_{\nu 0}(t)$ and $\rho_{\nu 1}(t)$ denote the probabilities of the empty (0) and
occupied (1) states, respectively, which satisfy the conservation of
probability $\rho_{\nu 0}(t)+\rho_{\nu 1}(t)=1$ for all time $t$.
The master equation is
\begin{align}
\frac{d}{dt}\vec{\rho}_{\nu}(t)&=\hat{M}_{\nu}\vec{\rho}_{\nu}(t),
\end{align}
where the transition rate matrix is
\begin{align}
\hat{M}_{\nu}&=
\left(
  \begin{array}{cc}
    -\Gamma_{\mathrm{l}\nu} & \Gamma_{\mathrm{u}\nu} \\
    \Gamma_{\mathrm{l}\nu} & -\Gamma_{\mathrm{u}\nu} 
  \end{array}
\right).
\end{align}
We are interested in a unique steady state, $\vec{\rho}_{\nu}^{\ \mathrm{st}}$, 
that satisfies the relation
\begin{align}
\left(
  \begin{array}{cc}
    -\Gamma_{\mathrm{l}\nu} & \Gamma_{\mathrm{u}\nu} \\
    1 & 1
  \end{array}
\right)\vec{\rho}_{\nu}^{\ \mathrm{st}}&=
\left(
  \begin{array}{c}
    0 \\
    1
  \end{array}
\right).
\end{align}
The solution is
\begin{align}
\rho_0^{\mathrm{st}}=\frac{\Gamma_{\mathrm{u}\nu}}{\Gamma_{\nu}},
\ \ & \rho_1^{\mathrm{st}}=\frac{\Gamma_{\mathrm{l}\nu}}{\Gamma_{\nu}},
\end{align}
where $\Gamma_{\nu}\equiv \Gamma_{\mathrm{l}\nu}+\Gamma_{\mathrm{u}\nu}$.
The steady current $I_{\nu}$ is evaluated by adding a counting field $i\chi$ to the
term representing the tunneling-out process to the upper electrode to the transition rate matrix: 
\begin{align}
\hat{M}_{\nu}(\chi)&=
\left(
  \begin{array}{cc}
    -\Gamma_{\mathrm{l}\nu} & \Gamma_{\mathrm{u}\nu}e^{i\chi} \\
    \Gamma_{\mathrm{l}\nu} & -\Gamma_{\mathrm{u}\nu} 
  \end{array}
\right),
\end{align}
and 
\begin{align}
I_{\nu}&=-e\frac{\partial}{\partial (i\chi)}\hat{M}_{\nu}(\chi)
\vec{\rho}_{\nu}^{\ \mathrm{st}}\notag\\
&=-e\left(
  \begin{array}{cc}
    0 & \Gamma_{\mathrm{u}\nu} \\
    0 & 0
  \end{array}
\right)
\frac{1}{\Gamma_{\nu}}
\left(
  \begin{array}{c}
    \Gamma_{\mathrm{u}\nu} \\
    \Gamma_{\mathrm{l}\nu}
  \end{array}
\right)\notag\\
&=-e\frac{\Gamma_{\mathrm{l}\nu}\Gamma_{\mathrm{u}\nu}}{\Gamma_{\nu}}.
\end{align}
The total steady current flowing from the lower to upper electrode is
\begin{align}
\left.I\right|_{\Gamma_{\mathrm{DD}}=0}&=-e\left\{
\frac{\Gamma_{\mathrm{lL}}\Gamma_{\mathrm{uL}}}{\Gamma_{L}}
+\frac{\Gamma_{\mathrm{lR}}\Gamma_{\mathrm{uR}}}{\Gamma_{R}}\right\}.
\end{align}

 \section{Coupled current path}
We now consider the effect of finite $\Gamma_{\mathrm{DD}}$.
 First we consider that an electron can tunnel only from dot L 
to dot R under a certain bias and gate voltage condition
 (as clarified in the discussion section).
 We define the probability vector of the two QD as
 \begin{align}
 \vec{\rho}(t)&=\left(
  \begin{array}{c}
    \rho_{00}(t) \\
    \rho_{10}(t) \\
    \rho_{01}(t) \\
    \rho_{11}(t)
  \end{array}
\right),
\end{align}
where $\rho_{nm}(t)$ denotes probability of $n\ (m)$ electrons in dot L (R).
The conservation of probability requires that 
$\sum_{n=0}^1\sum_{m=0}^1\rho_{nm}(t)=1$ for all time $t$.
 The master equation reads
 \begin{align}
 \frac{d}{dt}\vec{\rho}(t)&=\hat{M}\vec{\rho}(t),
 \end{align}
 where the transition rate matrix is
 \begin{align}
 \hat{M}&=\left(
  \begin{array}{cccc}
    -\Gamma_{\mathrm{lL}}-\Gamma_{\mathrm{lR}} & \Gamma_{\mathrm{uL}}
    & \Gamma_{\mathrm{uR}} & 0 \\
    \Gamma_{\mathrm{lL}} 
    & -\Gamma_{\mathrm{uL}}-\Gamma_{\mathrm{lR}}-\Gamma_{\mathrm{DD}} 
    & 0 & \Gamma_{\mathrm{uR}} \\
    \Gamma_{\mathrm{lR}} & \Gamma_{\mathrm{DD}} &
    -\Gamma_{\mathrm{uR}}-\Gamma_{\mathrm{lL}} & \Gamma_{\mathrm{uL}} \\
    0 & \Gamma_{\mathrm{lR}} & \Gamma_{\mathrm{lL}}
    &-\Gamma_{\mathrm{uR}}-\Gamma_{\mathrm{uL}}
  \end{array}
\right).
\end{align}
To find the steady state condition, we set 
\begin{align}
\left(
  \begin{array}{cccc}
    -\Gamma_{\mathrm{lL}}-\Gamma_{\mathrm{lR}} & \Gamma_{\mathrm{uL}}
    & \Gamma_{\mathrm{uR}} & 0 \\
    \Gamma_{\mathrm{lL}} 
    & -\Gamma_{\mathrm{uL}}-\Gamma_{\mathrm{lR}}-\Gamma_{\mathrm{DD}} 
    & 0 & \Gamma_{\mathrm{uR}} \\
    \Gamma_{\mathrm{lR}} & \Gamma_{\mathrm{DD}} &
    -\Gamma_{\mathrm{uR}}-\Gamma_{\mathrm{lL}} & \Gamma_{\mathrm{uL}} \\
    1 & 1 & 1 & 1
  \end{array}
\right)\vec{\rho}^{\ \mathrm{st}}&=
\left(
  \begin{array}{c}
    0 \\
    0 \\
    0 \\
    1
  \end{array}
\right).
\end{align}
The solution $\vec{\rho}^{\ \mathrm{st}}$ is obtained by inverting the matrix.
The steady current is derived as 
\begin{align}\label{eq:ltor}
I^{\mathrm{left}\to\mathrm{right}}&=-e\left(
  \begin{array}{cccc}
    0 & \Gamma_{\mathrm{uL}} & \Gamma_{\mathrm{uR}} & 0 \\
    0 & 0 & 0 & \Gamma_{\mathrm{uR}} \\
    0 & 0 & 0 & \Gamma_{\mathrm{uL}} \\
    0 & 0 & 0 & 0 
  \end{array}
\right)
\vec{\rho}^{\ \mathrm{st}}\notag\\
&=\left.I\right|_{\Gamma_{\mathrm{DD}}=0}-e
\frac{\Gamma_{\mathrm{uR}}\Gamma_{\mathrm{lL}}
(\Gamma_{\mathrm{uR}}\Gamma_{\mathrm{lL}}
-\Gamma_{\mathrm{uL}}\Gamma_{\mathrm{lR}})\Gamma_{\mathrm{DD}}}
{\Gamma_{\mathrm{L}}\Gamma_{\mathrm{R}}
\left\{\Gamma_{\mathrm{L}}\Gamma_{\mathrm{R}}+
\left(\Gamma_{\mathrm{uR}}
+\frac{(\Gamma_{\mathrm{lL}}+\Gamma_{\mathrm{uL}})
(\Gamma_{\mathrm{lL}}+\Gamma_{\mathrm{lR}})}
{\Gamma_{\mathrm{L}}+\Gamma_{\mathrm{R}}}
\right)\Gamma_{\mathrm{DD}}\right\}}.
\end{align}

In a similar calculation, we found the steady state current 
when the electron is allowed to tunnel only from dot R to dot L: 
\begin{align}\label{eq:rtol}
I^{\mathrm{right}\to\mathrm{left}}
&=\left.I\right|_{\Gamma_{\mathrm{DD}}=0}-e
\frac{\Gamma_{\mathrm{uL}}\Gamma_{\mathrm{lR}}
(\Gamma_{\mathrm{uL}}\Gamma_{\mathrm{lR}}
-\Gamma_{\mathrm{uR}}\Gamma_{\mathrm{lL}})\Gamma_{\mathrm{DD}}}
{\Gamma_{\mathrm{L}}\Gamma_{\mathrm{R}}
\left\{\Gamma_{\mathrm{L}}\Gamma_{\mathrm{R}}+
\left(\Gamma_{\mathrm{uL}}
+\frac{(\Gamma_{\mathrm{lR}}+\Gamma_{\mathrm{uR}})
(\Gamma_{\mathrm{lL}}+\Gamma_{\mathrm{lR}})}
{\Gamma_{\mathrm{L}}+\Gamma_{\mathrm{R}}}
\right)\Gamma_{\mathrm{DD}}\right\}}.
\end{align}

From these results, we see no effect of the interdot transition
when the relation $\Gamma_{\mathrm{uR}}/\Gamma_{\mathrm{lR}}
=\Gamma_{\mathrm{uL}}/\Gamma_{\mathrm{lL}}$ holds
which is usually satisfied under the experimental conditions discussed in our manuscript.
However, under a very strong magnetic field, when the wavefunctions 
in the electrode are strongly localized and are quite sensitive to the
fluctuations of the tunneling barrier thickness or impurities near
the tunneling barrier, the situations 
$\Gamma_{\mathrm{uR}}/\Gamma_{\mathrm{lR}}
\ll \Gamma_{\mathrm{uL}}/\Gamma_{\mathrm{lL}}$ 
(when $I^{\mathrm{left}\to\mathrm{right}}$ is suppressed
and $I^{\mathrm{right}\to\mathrm{left}}$ is enhanced)
or 
$\Gamma_{\mathrm{uR}}/\Gamma_{\mathrm{lR}}
\gg \Gamma_{\mathrm{uL}}/\Gamma_{\mathrm{lL}}$ 
(when $I^{\mathrm{left}\to\mathrm{right}}$ is enhanced
and $I^{\mathrm{right}\to\mathrm{left}}$ is suppressed)
can arise. The current is then enhanced or suppressed 
from that of the independent case $\left.I\right|_{\Gamma_{\mathrm{DD}}=0}$.

\section{Discussion}
\subsection{Stability diagram}
\begin{figure}[h]
\begin{center}\leavevmode
\includegraphics[width=0.7\linewidth]{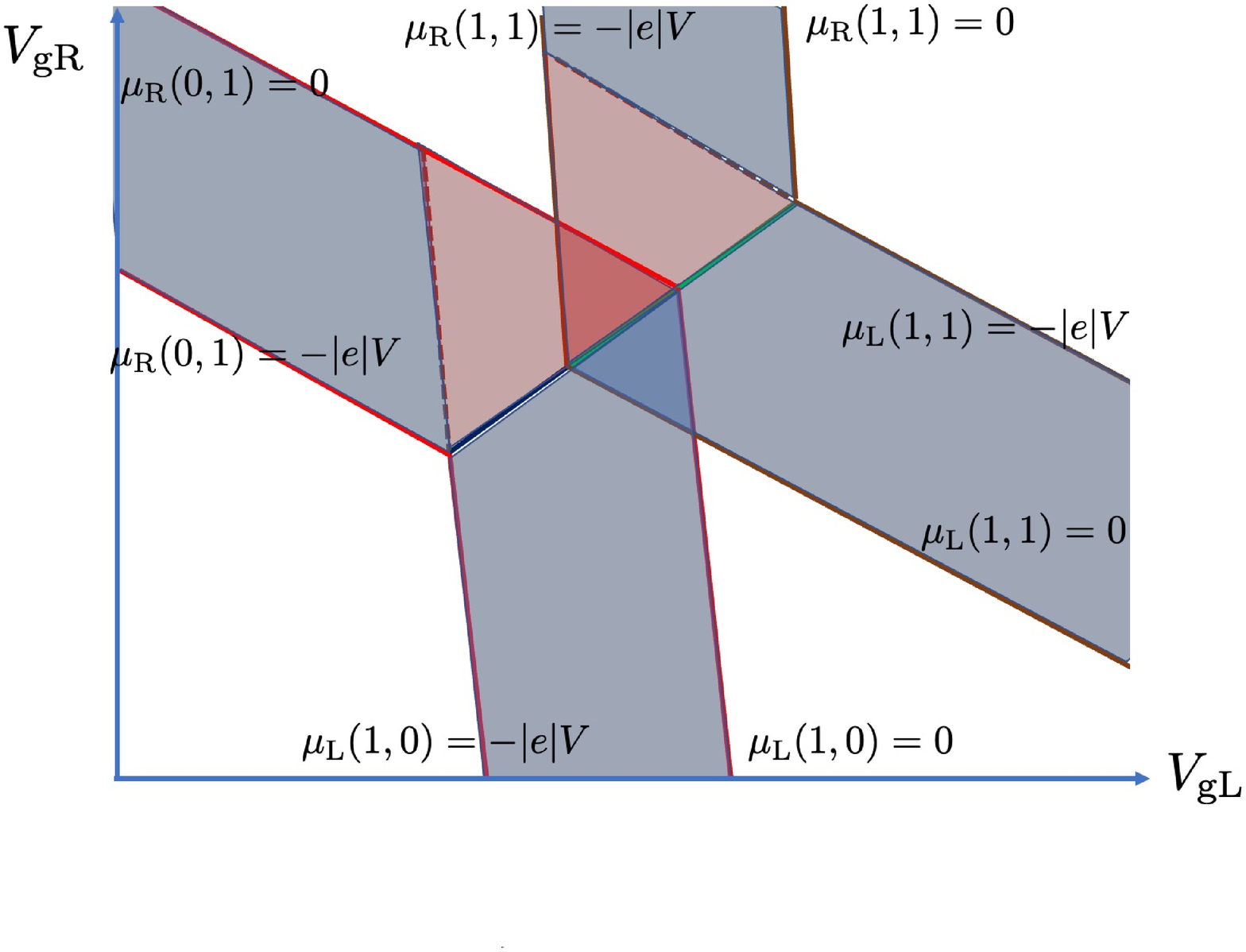}
\caption{Stability diagram when a finite bias $V_b$ is applied to the lower
electrode. 
In the red triangular regions, the current is enhanced in a polarity dependent manner.
} 
\label{fig:stability}
\end{center}
\end{figure}

The stability diagram of this system (see Fig.~\ref{fig:stability}) 
was constructed following the argument of \cite{wilfred}.
$\mu_{\mathrm{L/R}}(n,m)$ is the chemical potential of dot L or R
when there are $n\ (m)$ electrons in dot L (R).
The red triangular regions in the stability diagram indicate the regions of enhanced current
when tunneling from dot L to dot R is allowed, and
the condition 
$\Gamma_{\mathrm{uR}}/\Gamma_{\mathrm{lR}}
\gg \Gamma_{\mathrm{uL}}/\Gamma_{\mathrm{lL}}$ 
is satisfied.

\subsection{Bias polarity dependence}
If the polarity of the applied bias is changed while the other device conditions are fixed, 
we must exchange $\Gamma_{lR}$ and $\Gamma_{lL}$ with
$\Gamma_{uR}$ and $\Gamma_{uL}$, respectively.
Moreover, the interdot coupling rate $\Gamma_{\mathrm{DD}}$ can become 
$\Gamma_{\mathrm{DD}}'$.
When the line-widths are asymmetric, either of the currents given by Eqs.~(\ref{eq:ltor}) and (\ref{eq:rtol})
is enhanced.
Let us assume that the current of  Eq.~(\ref{eq:ltor}) is enhanced under the condition
$\Gamma_{\mathrm{uR}}/\Gamma_{\mathrm{lR}}
\gg \Gamma_{\mathrm{uL}}/\Gamma_{\mathrm{lL}}$.
By changing the polarity, the condition on the line-widths changes to
$\Gamma_{\mathrm{lR}}/\Gamma_{\mathrm{uR}}
\gg \Gamma_{\mathrm{lL}}/\Gamma_{\mathrm{uL}}$ and the current of Eq.~(\ref{eq:rtol}) is enhanced.
To compare the two enhanced currents, we replace the electrode indices in Eq.~(\ref{eq:rtol}) 
to get 
\begin{align}\label{eq:rtolpol}
I^{\mathrm{right}\to\mathrm{left}}_{\mathrm{reverse}}
&=\left.I\right|_{\Gamma_{\mathrm{DD}}=0}-e
\frac{\Gamma_{\mathrm{lL}}\Gamma_{\mathrm{uR}}
(\Gamma_{\mathrm{lL}}\Gamma_{\mathrm{uR}}
-\Gamma_{\mathrm{lR}}\Gamma_{\mathrm{uL}})\Gamma_{\mathrm{DD}}'}
{\Gamma_{\mathrm{L}}\Gamma_{\mathrm{R}}
\left\{\Gamma_{\mathrm{L}}\Gamma_{\mathrm{R}}+
\left(\Gamma_{\mathrm{lL}}
+\frac{(\Gamma_{\mathrm{uR}}+\Gamma_{\mathrm{lR}})
(\Gamma_{\mathrm{uL}}+\Gamma_{\mathrm{uR}})}
{\Gamma_{\mathrm{L}}+\Gamma_{\mathrm{R}}}
\right)\Gamma_{\mathrm{DD}}'\right\}}.
\end{align}
Comparing Eqs.~(\ref{eq:ltor}) and (\ref{eq:rtolpol}), 
the current values are the same if $\Gamma_{\mathrm{DD}}=\Gamma_{\mathrm{DD}}'$. 
To see this, note that 
$\Gamma_{\mathrm{uR}}(\Gamma_{\mathrm{L}}+\Gamma_{\mathrm{R}})
+(\Gamma_{\mathrm{lL}}+\Gamma_{\mathrm{uL}})(\Gamma_{\mathrm{lL}}+\Gamma_{\mathrm{lR}})
=\Gamma_{\mathrm{uL}}(\Gamma_{\mathrm{L}}+\Gamma_{\mathrm{R}})
+(\Gamma_{\mathrm{lR}}+\Gamma_{\mathrm{uR}})(\Gamma_{\mathrm{lL}}+\Gamma_{\mathrm{lR}})$.
Therefore, the current values in the triangles for one polarity should equal those in  
the same triangles for the reversed polarity.
However, the situation in our manuscript is rather more complicated 
because one of the QDs contains three or four electrons, whereas the other QD can accommodate 
at most one electron. 
Thus, the condition of $\Gamma_{\mathrm{DD}}\ne \Gamma_{\mathrm{DD}}'$ can be
realized in general.

\subsection{Pauli spin blockade}

It should be noted  that when the Pauli-spin blockade condition is satisfied
and the spin degree of freedom is properly included, part of the current-enhanced region disappears.
However, the blockade condition can be annulled 
in the parallel-coupled QD configuration, because an electron in the high spin state can  
escape to the upper electrode with a certain probability, although the rate is not high 
when the Pauli- spin blockade emerges.


\end{document}